\documentclass{article}
\usepackage{amsmath, amssymb, amsfonts}
\title{ Screening an extremal black hole with a thin shell of exotic matter } 
\author{Hristu Culetu, \\Ovidius University, Department of Physics and Electronics, \\ Mamaia Avenue 124, 900527 Constanta, Romania, \\e-mail : hculetu@yahoo.com}

\begin{document}
\numberwithin{equation}{section}
\pagenumbering{arabic}
\maketitle
\newcommand{\fv}{\boldsymbol{f}}
\newcommand{\tv}{\boldsymbol{t}}
\newcommand{\gv}{\boldsymbol{g}}
\newcommand{\OV}{\boldsymbol{O}}
\newcommand{\wv}{\boldsymbol{w}}
\newcommand{\WV}{\boldsymbol{W}}
\newcommand{\NV}{\boldsymbol{N}}
\newcommand{\hv}{\boldsymbol{h}}
\newcommand{\yv}{\boldsymbol{y}}
\newcommand{\RE}{\textrm{Re}}
\newcommand{\IM}{\textrm{Im}}
\newcommand{\rot}{\textrm{rot}}
\newcommand{\dv}{\boldsymbol{d}}
\newcommand{\grad}{\textrm{grad}}
\newcommand{\Tr}{\textrm{Tr}}
\newcommand{\ua}{\uparrow}
\newcommand{\da}{\downarrow}
\newcommand{\ct}{\textrm{const}}
\newcommand{\xv}{\boldsymbol{x}}
\newcommand{\mv}{\boldsymbol{m}}
\newcommand{\rv}{\boldsymbol{r}}
\newcommand{\kv}{\boldsymbol{k}}
\newcommand{\VE}{\boldsymbol{V}}
\newcommand{\sv}{\boldsymbol{s}}
\newcommand{\RV}{\boldsymbol{R}}
\newcommand{\pv}{\boldsymbol{p}}
\newcommand{\PV}{\boldsymbol{P}}
\newcommand{\EV}{\boldsymbol{E}}
\newcommand{\DV}{\boldsymbol{D}}
\newcommand{\BV}{\boldsymbol{B}}
\newcommand{\HV}{\boldsymbol{H}}
\newcommand{\MV}{\boldsymbol{M}}
\newcommand{\be}{\begin{equation}}
\newcommand{\ee}{\end{equation}}
\newcommand{\ba}{\begin{eqnarray}}
\newcommand{\ea}{\end{eqnarray}}
\newcommand{\bq}{\begin{eqnarray*}}
\newcommand{\eq}{\end{eqnarray*}}
\newcommand{\pa}{\partial}
\newcommand{\f}{\frac}
\newcommand{\FV}{\boldsymbol{F}}
\newcommand{\ve}{\boldsymbol{v}}
\newcommand{\AV}{\boldsymbol{A}}
\newcommand{\jv}{\boldsymbol{j}}
\newcommand{\LV}{\boldsymbol{L}}
\newcommand{\SV}{\boldsymbol{S}}
\newcommand{\av}{\boldsymbol{a}}
\newcommand{\qv}{\boldsymbol{q}}
\newcommand{\QV}{\boldsymbol{Q}}
\newcommand{\ev}{\boldsymbol{e}}
\newcommand{\uv}{\boldsymbol{u}}
\newcommand{\KV}{\boldsymbol{K}}
\newcommand{\ro}{\boldsymbol{\rho}}
\newcommand{\si}{\boldsymbol{\sigma}}
\newcommand{\thv}{\boldsymbol{\theta}}
\newcommand{\bv}{\boldsymbol{b}}
\newcommand{\JV}{\boldsymbol{J}}
\newcommand{\nv}{\boldsymbol{n}}
\newcommand{\lv}{\boldsymbol{l}}
\newcommand{\om}{\boldsymbol{\omega}}
\newcommand{\Om}{\boldsymbol{\Omega}}
\newcommand{\Piv}{\boldsymbol{\Pi}}
\newcommand{\UV}{\boldsymbol{U}}
\newcommand{\iv}{\boldsymbol{i}}
\newcommand{\nuv}{\boldsymbol{\nu}}
\newcommand{\muv}{\boldsymbol{\mu}}
\newcommand{\lm}{\boldsymbol{\lambda}}
\newcommand{\Lm}{\boldsymbol{\Lambda}}
\newcommand{\opsi}{\overline{\psi}}
\renewcommand{\tan}{\textrm{tg}}
\renewcommand{\cot}{\textrm{ctg}}
\renewcommand{\sinh}{\textrm{sh}}
\renewcommand{\cosh}{\textrm{ch}}
\renewcommand{\tanh}{\textrm{th}}
\renewcommand{\coth}{\textrm{cth}}

\begin{abstract}
We study the possibility of shielding a regular extremal black hole by means of a matter thin-shell. While the surface energy density $\sigma$ on the static shell is negative, the tangential pressures $p$ are positive, both of them being finite when the shell approaches the black hole horizon. The Darmois-Israel junction conditions are used to find $\sigma$ and $p$ in terms of the radius $a$ of the shell. The surface gravitational energy $E_{S}$ is computed, keeping track of the pressure contribution. The stability conditions are briefly investigated.
 \end{abstract}
 
\section{Introduction}
 Once the surface-layer formalism has been developed, one naturally asks the question whether the black hole (BH) hairs could be shielded against exterior observers \cite{PV, SG, MH}. Eiroa and Simeone \cite{ES1} investigated the stability of charged thin-shells (see also \cite{ES2}) whereas Pereira, Coelho and Rueda \cite{PCR} studied the stability of thin-shell interfaces inside compact stars. More recently, Mazharimousavi and Halilsoy (MH) \cite{MH} applied the Israel junction conditions on a dust thin shell that separates an inner Reissner-Nordstrom (RN) black hole from an outer Schwarzschild (KS) geometry. They showed that the electric charge of the RN black hole may be screened for some value of the KS mass when the boundary conditions at the thin shell are satisfied. Their shell is like a bubble of exotic matter due to the negative surface energy density. In addition, MH stated that, when the RN black hole of mass $m$ is extremal, the exterior KS mass should vanish and the shell energy becomes $-m$. 

  Our purpose in this paper concerns the study of a regular BH \cite{HC1, HC2} surrounded by a spherical thin shell which is capable of screening the gravitational field of the BH (namely, outside the shell the metric is Minkowski), for certain values of the parameters of the physical system. In contrast with MH, we endow the static shell with a positive pressure for to get flat space outside it. For the junction relations to be fulfilled, the surface energy density has to be negative. Moreover, a finite solution is obtained when the shell radius approaches the KS radius of the BH, with a Minkowskian geometry outside the shell (total screening of the BH - as if it were dark matter). 
	
	We use geometrical units $G = c = 1$ throughout the paper.
	
	\section{ Black hole screening}
 To avoid any singularity in the line element we consider a modified Schwarzschild BH is located at the origin of coordinates \cite{HC1, HC2}
  \begin{equation}
   ds_{-}^{2} = -\left(1 - \frac{2m}{r} e^{-\frac{r_{H}}{r}}\right) dt^{2} + \frac{1}{1 - \frac{2m}{r} e^{-\frac{r_{H}}{r}}} dr^{2} + r^{2} d \Omega^{2},     
 \label{2.1}
 \end{equation} 
wher $m$ is the BH mass, $d \Omega^{2}$ stands for the metric on the unit 2-sphere and $r_{H} = 2m/e$ is the event horizon radius, with $lne = 1$. (2.1) is an extremal BH, with a vanishing surface gravity \cite{HC1} and with positive metric coefficients. It is a modified version of the Xiang et al \cite{XLS} spacetime, where $-g_{tt} = 1/g_{rr} = 1 - (2m/r)exp(-\alpha/r^{2})$, $\alpha$ being a length squared. To reach (2.1) we noticed in \cite{HC4} that, at large distances, their metric does not retrieve the Reissner-Nordstrom or the extremal BH geometries when a second order radial power expansion of $g_{tt}$ is performed: $-g_{tt} \approx 1-2m/r+2\alpha m/r^{3}$, when $\alpha$ is related to the BH mass or charge. Therefore, we consider (2.1) a better choice than that of Xiang et al., for to get rid of the KS singularity. 

We surround the BH with a spherical spacelike thin shell located at $r = a >r_{H},~a = const.$. The static shell is endowed with a surface energy density $\sigma$ and tangential surface pressures $p_{\theta} = p_{\phi} = p$. The surface stress tensor corresponds to a perfect fluid 
   \begin{equation}
   S^{i}_{j} = diag(-\sigma, p, p), ~~~i,j = \tau, \theta, \phi
 \label{2.2}
 \end{equation} 
and the metric on the shell $\Sigma$ is given by
   \begin{equation}
    ds_{\Sigma}^{2} = -d\tau^{2} + a^{2} d \Omega^{2},  
 \label{2.3}
 \end{equation} 
where $\tau$ is the proper time of an observer on the shell. Our aim is to find $\sigma(a)$ and $p(a)$ such that $\Sigma$ would play the role of a total absorber, namely the BH is shielded and the metric outside the shell of matter is Minkowskian
  \begin{equation}
  ds_{+}^{2} = -dt^{2} + dr^{2} + r^{2} d \Omega^{2}.
 \label{2.4}
 \end{equation} 
To reach that purpose we must impose the Gauss-Codazzi boundary conditions at $r = a$. We are not going to repeat here the details of the formalism (see, for example, \cite{GLV, HC3}. We write down only the Lanczos equation
  \begin{equation}
   -8\pi S^{i}_{~j} = [K^{i}_{~j}] - \delta^{i}_{~j}[K^{l}_{l}],
 \label{2.5}
 \end{equation} 
where $[K^{i}_{~j}] = K^{i}_{~j}(out) - K^{i}_{~j}(in)$ is the jump of the second fundamental form when the shell is crossed and $[K^{l}_{~l}]$ is the jump of the mean extrinsic curvature. For our static shell, one obtains \cite{MH, GLV}
   \begin{equation}
   4\pi a \sigma = \sqrt{f} - 1
 \label{2.6}
 \end{equation} 
and
   \begin{equation}
  8\pi p = \frac{1 - \sqrt{f}}{a} - \frac{f'}{2\sqrt{f}},
 \label{2.7}
 \end{equation} 
with $f = 1 - (2m/r) e^{-r_{H}/r},~f' = df/dr$ and all functions are calculated at $r = a$. Keeping in mind that $0 < f < 1$, we have $0 > \sigma >-1/4\pi a$. We may, in principle, determine the direct relation between $\sigma$ and $p$ (namely, $p = p(\sigma)$ - the equation of state of the fluid) once $a$ and $m$ are given. We also notice that the energy conditions for $S^{i}_{j}$ are not obeyed, as is usually valid for exotic matter.

From (2.6) and (2.7) the following equation can be obtained
   \begin{equation}
  4\pi (\sigma + 2p) = - \frac{m(1 - \frac{r_{H}}{a})e^{-\frac{r_{H}}{a}}}{a^{2} \sqrt{1 - \frac{2m}{a} e^{-\frac{r_{H}}{a}}}},~~~a>r_{H}.   
 \label{2.8}
 \end{equation} 
We recognize at the r.h.s. of the above equation the (minus) proper acceleration $\sqrt{ a^{b}a_{b}}$ of a static observer in the geometry (2.1) \cite{HC1}, i.e. $4\pi (\sigma + 2p) = - \sqrt{ a^{b}a_{b}}$, with $a^{b} = u^{a}\nabla_{a}u^{b}, u^{b} = (1/\sqrt{f(a)}, 0, 0, 0)$ and $a, b$ span the coordinates $t, r, \theta, \phi$. We may also write
   \begin{equation}
  4\pi (\sigma + 2p) = - a^{b}n_{b}  
 \label{2.9}
 \end{equation} 
where $n_{b} = (0, 1/\sqrt{f(a)}, 0, 0)$ is the normal vector to the $r = a$ hypersurface \cite{MGLV}. From (2.6) one observes that $\sigma$ is always negative, $\sigma(a) < 0$, because $0 < f(a) <1$ \cite{HC1}. In addition, $\sigma(a)$ becomes zero when $a \rightarrow \infty$ and attains its minimal value $-1/4\pi r_{H} = -e/8\pi m$ when $a \rightarrow r_{H}$. This can be seen from the expression of $\sigma'(a) \equiv d\sigma/da$
   \begin{equation}
  \sigma'(a) =  \frac{1 - \sqrt{f(a)}}{4\pi a^{2}} + \frac{f'(a)}{8\pi a\sqrt{f(a)}} > 0,
 \label{2.10}
 \end{equation} 
because $f(a) < 1$ and 
   \begin{equation}
  f'(a) = \frac{2m(1 - \frac{r_{H}}{a})e^{-\frac{r_{H}}{a}}}{a^{2}} >0
 \label{2.11}
 \end{equation} 
for $a > r_{H}$.
Note that the pressure $p$ from (2.7) may be written in terms of $\sigma(a)$ and its derivative
   \begin{equation}
   p = -\sigma - \frac{a\sigma'(a)}{2},
 \label{2.12}
 \end{equation} 
 which is just the conservation equation \cite{MH, ES1}.
 
It is worth observing that, for $a >> r_{H}$, i.e. when $f \approx 1$ and the exponential factor is taken equal to unity, $\sigma$ acquires the form
   \begin{equation}
 \sigma = -\frac{m e^{-\frac{r_{H}}{a}}}{2\pi a^{2} (1 + \sqrt{f(a)})} \approx -\frac{m}{4\pi a^{2}}  
 \label{2.13}
 \end{equation} 
but $p$ is vanishing ($r_{H}/a = 2m/ea$ is neglected w.r.t. unity). A similar result for $\sigma$ has been obtained by the authors of \cite{MH} when the BH is extremal. However, in our situation (2.13) is valid only asymptotically. We notice that when $a \rightarrow r_{H}, ~f \rightarrow 0$ and $p$ is undetermined (one obtains $0/0$). To find the limit, we expand $f(a)$  around $a = r_{H}$ \cite{HC1}, to second order in $a - r_{H}$,
   \begin{equation}
   f(a) \approx f(r_{H}) + (a - r_{H}) f'(r_{H}) + \frac{1}{2} (a - r_{H})^{2} f''(r_{H}) = \frac{(a - r_{H})^{2}}{2r_{H}^{2}}
 \label{2.14}
 \end{equation} 
with the result $p \rightarrow e(2 - \sqrt{2})/32\pi m >0 $. In contrast to the KS metric, the second term in the r.h.s. of (2.7) is finite when $a \rightarrow r_{H}$ and $p$ is not divergent there (see \cite{AL1, AL2} where the authors emphasize the role of the singularity of the invariant acceleration on the KS horizon). It should also be remarked that, if we chose a KS black hole instead of our modified form, $p$ would have been divergent at $a = r_{H}$, too.              

\section{Thin shell energy}
Having a negative $\sigma$, our shell is like a bubble of exotic matter. We look for the active gravitational energy $E_{S}$ \cite{SG, MGLV, HC1} on this hypersurface. As Martin-Moruno et al. \cite{MGLV} have noticed, the combination $4\pi (\sigma + 2p)$, which is the discontinuity of the 4-acceleration for a thin shell, has similar properties to those of the quantity $\rho +3p$ for the bulk spacetime (see also \cite{TP}). Therefore, to find $E_{S}$ we keep track of the pressure contribution and write
    \begin{equation}
		E_{S} = \int^{\infty}_{0} \int^{\pi}_{0} \int^{2\pi}_{0} (\sigma +2p)\delta(r-a)\sqrt{-g}dr d\theta d\phi
 \label{3.1}
 \end{equation} 
which yields
    \begin{equation}
		E_{S} = 4\pi a^{2} (\sigma + 2p)= -a^{2} (a^{b}n_{b}) = -a^{2} \frac{f'(a)}{2\sqrt{f(a)}} < 0.
 \label{3.2}
 \end{equation} 
Let us investigate the behaviour of $E_{S}$ when the shell approaches the BH horizon ($a \rightarrow r_{H}$). From (2.8) we conclude that $\sigma +2p$ is not defined at $a = r_{H}$. However, we may consider the limit $a \rightarrow r_{H} (a > r_{H}$) and obtain
 \begin{equation}
(\sigma +2p)_{H} = -\frac{\sqrt{2}}{8\pi r_{H}} = -\frac{e\sqrt{2}}{16\pi m},
 \label{3.3} 
 \end{equation} 
whence $E_{S} = (\sigma +2p)_{H} \cdot 4\pi r_{H}^{2} = -\sqrt{2}m/e$. 

To summarize, our system is composed by a BH whose horizon turns out to be endowed with a negative energy $E_{S}$ because of which the BH is screened, becoming gravitationally invisible to the outside world, which is Minkowskian. We remind that our BH is extremal and, therefore, its Hawking temperature is vanishing \cite{HC1}. Note also that the surface energy from MH model \cite{MH} (their energy $\Omega$) vanishes at $a = r_{e}$ (their event horizon) because their surface energy density $\sigma$ becomes zero. Hence, there is no any MH thin shell of exotic dust matter , very close to the horizon.

\section{Stability considerations}
As we have shown, we found a thin-shell of exotic matter surrounding an extremal BH so that the outcome for an external observer is a flat spacetime. An important problem concerns the mechanical stability of such a hypersurface of exotic matter. 

 We now consider small spherical perturbations around the static solution of the equations above. Let the radius of the shell to depend on the proper time $\tau$ on the shell. In that case, Eq. (2.6) can be written as \cite{MH, ES1}
    \begin{equation}
   4\pi a \sigma = \sqrt{f + \dot{a}^{2}} - \sqrt{1 + \dot{a}^{2}},
 \label{4.1}
 \end{equation} 
 with $\dot{a} = da/d\tau$. Squaring (4.1) twice, one obtains
     \begin{equation}
    \dot{a}^{2} + V(a) = 0
 \label{4.2}
 \end{equation} 
with
     \begin{equation}
    V(a) = \frac{f + 1}{2} - 4\pi^{2} \sigma^{2}a^{2} - \frac{(f - 1)^{2}}{64\pi^{2} \sigma^{2}a^{2}},
 \label{4.3}
 \end{equation} 
where $V(a)$ has the meaning of a potential, by analogy between (4.2) and the energy of a point particle with one degree of freedom \cite{ES1}. With $f$ from Ref.7 (see also below (2.7)), Eq. (4.3) becomes 
  \begin{equation}
    V(a) = 1 - \left(2\pi \sigma a + \frac{m}{4\pi \sigma a^{2}}e^{-\frac{r_{H}}{a}}\right)^{2}
 \label{4.4}
 \end{equation} 
The stability requires $V(a_{0}) = 0,~V'(a_{0})$ and $V''(a_{0}) < 0$ (because $\sigma < 0)$ \cite{PCR} at a radius $a = a_{0}$ for which $\dot{a} = 0$ (the prime denotes the derivative w.r.t. $a_{0}$). From (4.4) we have
  \begin{equation}
    V(a_{0}) = (1 - 2\pi \sigma a_{0} - \frac{m}{4\pi \sigma a_{0}^{2}}e^{-\frac{r_{H}}{a_{0}}}) (1 + 2\pi \sigma a_{0} + \frac{m}{4\pi \sigma a_{0}^{2}}e^{-\frac{r_{H}}{a_{0}}}) = 0.
 \label{4.5}
 \end{equation} 
The first factor on the r.h.s. of (4.5) is positive since $\sigma < 0$ and therefore one obtains
  \begin{equation}
     (1 + 2\pi \sigma a_{0} + \frac{m}{4\pi \sigma a_{0}^{2}}e^{-\frac{r_{H}}{a_{0}}}) = 0.
 \label{4.6}
 \end{equation} 
One sees that (4.6) has a solution for $\sigma ({a_{0}})$ only when $1 + 2\pi \sigma a_{0} > 0$, or $\sigma > -1/2\pi a_{0}$. That is always valid from (2.6) because $0 < f <1$ for $a_{0} > r_{H}$.

\section{Conclusions}
 Our aim in this paper was to apply the Gauss-Codazzi equations to find whether is it possible to hide/screen an extremal BH by means of a spherical thin-shell of matter, surrounding the BH at a distance $a > r_{H}$. We found $\sigma$ and $p$ as functions of its radius $a > r_{H}$ with the help of the junction conditions. $\sigma(a)$ is a monotonic function and reaches its minimum $-e/8\pi m$ at $a = r_{H}$, becoming $-m/4\pi a^{2}$ at large distances w.r.t. $r_{H}$. In calculating the total shell energy we took into consideration the contribution from the tangential pressures, giving $E_{S} \rightarrow -\sqrt{2}m/e$ when the shell approaches the BH horizon. We also briefly studied the stability conditions of the system under investigation.\\
 
 \textbf{Acknowledgements}
 
  I am grateful to the anonymous referee for useful suggestions which improved the quality of the manuscript.

\end{document}